\documentclass[12pt,a4paper]{article}
\usepackage[leqno]{amsmath}
\title{Solitary wave solutions of several nonlinear PDEs 
modeling shallow water waves}
\author{Nikolay K. Vitanov, Tsvetelina I. Ivanova}
\date{Institute of Mechanics, Bulgarian Academy of Sciences, 
Akad. G. Bonchev Str., Bl. 4, 1113 Sofia, Bulgaria}
\begin{document}
\maketitle
\begin{abstract}
We apply the version of the method of simplest equation called modified method of simplest equation for obtaining exact traveling wave solutions of a class of equations that contain as particular case a nonlinear PDE that models shallow water waves in viscous fluid (Topper-Kawahara equation). As simplest equation we use a version of the Riccati equation. We obtain two exact traveling wave solutions of equations from the studied class of equations and discuss the question of imposing boundary conditions on one of these solutions.
\end{abstract}
\section{Introduction}
Nonlinear phenomena are of large interest for modern natural and
social sciences \cite{np1} - \cite{np5}. And many nonlinear phenomena are modelled by nonlinear differential equations \cite{n1} - \cite{debn}.
Today the research on nonlinear phenomena is well established research area
and the research on nonlinear waves finds its significant place within this
area. One of the most studied topics in the research on nonlinear phenomena are
the nonlinear waves and especially the water waves \cite{w1} -\cite{w4}.
The goal of obtaining exact solution of model equation for water waves
leaded to development of the modified method of simplest equation that will
be applied below in the text \cite{v1}-\cite{v3}.
Often the model equations for the nonlinear phenomena are nonlinear partial differential equations that have traveling-wave solutions. These traveling wave 
solutions are studied very intensively  \cite{scott}-\cite{ablowitz1}.
Well established methods exist for obtaining exact traveling-wave solutions
of integrable nonlinear partial differential equations 
, e.g., the method of inverse scattering transform or the method of Hirota 
\cite{gardner} - \cite{hirota}. Many other approaches for obtaining exact special solutions of nonlintegrable nonlinear PDEs have been developed in the recent years (for examples see \cite{he} - \cite{wazw2}). Below we shall consider
the method of simplest equation  and our focus will be on a version of this
method called modified method of simplest equation.
The method of simplest equation is based on  a procedure analogous to the first 
step of the test for the Painleve property \cite{kudr05x}, \cite{kudr08}. In the version of the method called  modified method of simplest equation \cite{v1} - \cite{v3} this procedure is substituted 
by  the concept for the balance equations.  
The modified method of simplest equation has already numerous applications,
e.g., obtaining exact traveling wave solutions  of generalized Kuramoto - Sivashinsky equation, reaction - diffusion  equation, reaction - telegraph equation\cite{v1}, \cite{vd10},
generalized Swift - Hohenberg equation and generalized Rayleigh equation \cite{v2}, 
generalized Fisher equation, generalized Huxley equation \cite{vit09x}, 
generalized Degasperis - Procesi equation and b-equation \cite{vit11a}, 
extended Korteweg-de Vries equation \cite{v13}, etc. \cite{v13a} - \cite{vit17}.
\par
The text below is organized as follows. In Sect. 2 we describe the methodology and formulate the problem. In Sect. 3 we present solutions of the studied nonlinear PDEs. Discussion and concluding remarks are given in Sect.4
\section{Methodology and problem formulation}
We shall use the part of the methodology of the modified method of simplest equation that is appropriate for solving model nonlinear PDEs for shallow water waves. This methodology  is described, e.g., in \cite{vit11b}. The methodology works as follows. First of all by means of an 
appropriate ansatz the solved  nonlinear partial differential 
equation is reduced to 
a nonlinear  differential equation, containing derivatives of a function
\begin{equation}\label{ix1}
P \left( u(\xi), u_{\xi},u_{\xi \xi},\dots,\right) = 0
\end{equation}
Then the function $u(\xi)$   is searched as some 
function of another function, e.g., $g(\xi)$, etc., i.e.
\begin{equation}\label{ix1x}
u(\xi) = G[g(\xi)]
\end{equation} 
The kind of the function $F$ is not
prescribed. Often one uses a finite-series relationship, e.g., 
\begin{equation}\label{i2}
u(\xi) = \sum_{\mu_1=-\nu_1}^{\nu_2} p_{\mu_1} [g (\xi)]^{\mu_1}; 
\end{equation}
where $p_{\mu_1}$ are coefficients (such relationship will be used below too). The function  $g(\xi)$ is
solution of simpler ordinary differential equation called simplest 
equation. Eq.(\ref{ix1x}) is substituted in Eq.(\ref{ix1}) and let the result
of this substitution be a polynomial containing $g(\xi)$. 
Next a balance procedure is applied.  This procedure has to ensure 
that all of the coefficients of the obtained polynomial of $g(\xi)$ 
contain more than one term. The procedure leads to a
balance equation for some of the parameters of the solved equation and
for some of the parameters of the solution. Eq.(\ref{ix1x}) describes a
candidate for solution of Eq.(\ref{ix1}) if all coefficients of the obtained 
polynomial of are equal to $0$. This condition leads to a system of 
nonlinear algebraic equations for the coefficients of the solved nonlinear PDE and for the
coefficients of the solution. Any nontrivial solution of this algebraic system
leads to a solution the studied  nonlinear partial differential equation.
\par
We shall apply the above methodology to the class of equations
\begin{equation}\label{class}
\frac{\partial u}{\partial t} + u \frac{\partial u}{\partial x} + r \frac{\partial^2 u}{\partial x^2} +
\delta \frac{\partial^3 u}{\partial x^3} + \sigma u^n \frac{\partial u^m}{\partial x^m} =0
\end{equation}
where $\sigma$, $r$ and $\delta$ are parameters and $n$ and $m$ are non-negative
natural numbers. For the case $n=0$, $m=4$, Eq.(\ref{class}) is reduced to an equation  modeling long nonlinear waves in viscous fluid (Topper - Kawahara
equation) \cite{top}. We shall search a traveling wave solution of Eq.(\ref{class}) in the form
\begin{equation}\label{s1}
u(\xi) = \sum \limits_{\mu=0}^\nu p_\mu g(\xi)^\mu
\end{equation}
where $\nu$ and $p_\mu$ are parameters and $\xi = \alpha x + \beta t$. The simplest equation for $g(\xi)$ will be
\begin{equation}\label{s2}
g' = \frac{dg}{d \xi} =\sum \limits_{\theta = 0}^\Theta q_\theta g^\theta
\end{equation}
where $q_\theta$ is parameter.
Next step is to obtain the possible balance equations by balancing the
powers of $g$ arising from the terms from Eq.(\ref{class}). The maximum powers of $g$ arising from the 5 terms in Eq.(\ref{class}) are
\begin{eqnarray*}
{\rm Term} \ \ \frac{\partial u}{\partial t} \to \nu + \Theta - 1; \ \
{\rm Term} \ \  u \frac{\partial u}{\partial x} \to 2 \nu + \Theta -1; \\
{\rm Term} \ \ r \frac{\partial^2 u}{\partial x^2} \to \nu + 2 (\Theta -1); \ \
{\rm Term} \ \ \delta \frac{\partial^3 u}{\partial x^3} \to \nu + 3(\Theta -1); 
\\ 
{\rm Term} \ \ \alpha u^n \frac{\partial u^m}{\partial x^m} \to n \nu + m(\Theta -1)
\end{eqnarray*}
Thus the possible members of a balance equation are
$$
A = 2 \nu + (\Theta -1); \ \ B = \nu + 3(\Theta -1); \ \ C = (n+1) \nu + m(\Theta -1)
$$
Let us consider the possible balance equations from the point of view of
increasing values of the parameter $m$. We have the following possibilities
\begin{enumerate}
\item $m=1$. The interesting cases are $n=0$ or $n \ge 2$.
\begin{enumerate}
\item $n=0$. The balance equation is $A=B$. Thus $\nu = 2 (\Theta -1)$.
\item $n \ge 2$. The balance equation is $B = C$. Thus $\nu = 2 \left(\frac{\Theta - 1}{n} \right)$.
\end{enumerate}
\item $m=2$. The interesting cases are: $n=1$ and $n >1$
\begin{enumerate}
\item $n=1$. The balance equation is $B=C$. Thus $\nu =  (\Theta -1)$. 
\item $n>1$. The balance equation is $B = C$. Thus $\nu = \left(\frac{\Theta - 1}{n} \right)$.
\end{enumerate}
\item $m \ge 3$. The interesting case is $n \ge 0$.
\begin{enumerate}
\item $n\ge 0$. The balance equation is $A=C$. Thus $\nu = \left( \frac{m-1}{1-n}\right)(\Theta -1)$.
\end{enumerate}
\end{enumerate}
Below we shall consider examples from the case $m=4$ and $m=2$.
\section{Several solutions of equations of class (\ref{class})}
Let us first discuss the case $m=4$. As we can see from the above list we must have $n=0$ and then
the balance equation becomes $\nu = 3 (\Theta -1)$. Let us discuss the case $\Theta=2$. Then $\nu=3$.
All above means that we discuss the equation
\begin{equation}\label{class1}
\frac{\partial u}{\partial t} + u \frac{\partial u}{\partial x} + r \frac{\partial^2 u}{\partial x^2} +
\delta \frac{\partial^3 u}{\partial x^3} + \sigma  \frac{\partial u^4}{\partial x^4} =0
\end{equation}
and search for the solution of the kind

\begin{equation}\label{y1}
u(\xi) = p_0 + p_1g(\xi) + p_2 g(\xi)^2 + p_3 g(\xi)^3
\end{equation}
where the simplest equation for $g(\xi)$ is
\begin{equation}\label{y2}
g'=q_0 + q_1 g + q_2 g^2
\end{equation}
We substitute Eqs. (\ref{y1}) and (\ref{y2}) in Eq.(\ref{class1}) and obtain the following
equation
\begin{equation} \label{y3}
A_0 + A_1 g(\xi) + A_2 g(\xi)^2 + A_3 g(\xi)^3 + A_4 g(\xi)^4 + A_5 g(\xi)^5 + A_6 g(\xi)^6 +
A_7 g(\xi)^7 =0
\end{equation}
where $A_i$, $i=0,\dots,7$, are nonlinear algebraic relationship between parameters of the equation and parameters of the solution. Further we set $A_i=0$ and obtain, e.g.,  the following solution of 
the system of $8$ nonlinear algebraic equations
\begin{eqnarray}\label{sol}
\sigma&=& \frac{\delta^2}{16 r} \nonumber \\ 
q_1 &=& 2 \frac{(q_0 q_2 \alpha^2 \delta^2 + 4 r^2)^{1/2}}{\alpha \delta} \nonumber \\
p_3&=&-\frac{15}{2} \frac{\delta^2 \alpha^2 q_2^3}{r} \nonumber \\
p_2&=& -\frac{15}{2} \alpha q_2^2 \delta \left[\frac{3 }{r}(q_0 q_2 \alpha^2 \delta^2 + 4r^2)^{1/2} +2 \right]\nonumber \\
p_1&=& - \frac{15}{2} \frac{q_2}{ r} \Bigg[3 q_0 q_2 \alpha^2 \delta^2 + 8  r^2 + 4 r (q_0 q_2 \alpha^2 \delta^2 + 4r^2)^{1/2} \Bigg]\nonumber \\
p_0&=&-\frac{1}{2 \alpha \delta} \Bigg[2 \beta \delta + 30 q_0 q_2 \alpha^2 \delta^2 + 48 r^2 + 15 q_0 q_2 \frac{\delta^2}{r} (q_0 q_2 \alpha^2 \delta^2 + 4r^2)^{1/2}\Bigg]\nonumber \\
\end{eqnarray}
Eq.(\ref{y2}) is a particular case of equation of Riccati. One particular
solution of the Riccati equation is 
\begin{equation}\label{y4}
g(\xi) = - \frac{q_1}{2 q_2} - \frac{\pi}{2 q_2} \tanh \left[ \frac{\pi}{2}(\xi + \xi_0) \right]
\end{equation}
where $\xi_0$ is a constant of integration and $\pi$ has to satisfy
$\pi^2 = q_1^2 - 4 q_0 q_2 >0$. The substitution of $q_1$ from Eq.(\ref{sol})
in the last inequality leads to $q_1^2 = 16 r^2/(\alpha^2 \delta^2)$ which is larger than
$0$ for nonzero values of the real parameters $r$ and $\delta$. Let us set
$\pi = 4r/(\alpha \delta)$. Then the solution (\ref{y4}) of Eq.(\ref{sol}) becomes
\begin{equation}\label{y5}
g(\xi) = - \frac{1}{q_2 \alpha \delta} \left \{  \left( q_0 q_2 \alpha^2 \delta^2 + 4r^2\right)^{1/2} + 2r \tanh \left[\frac{2r}{\alpha \delta} (\xi + \xi_0) \right]\right \}
\end{equation}
The substitution of Eq.(\ref{y5}) in Eq.(\ref{y1}) leads to the following
solution of Eq.(\ref{class1})
\begin{eqnarray}\label{y6}
u(\xi) = -\frac{1}{2 \alpha \delta} \Bigg[2 \beta \delta + 30 q_0 q_2 \alpha^2 \delta^2 + 48 r^2 + 15 q_0 q_2 \frac{\delta^2}{r} (q_0 q_2 \alpha^2 \delta^2 + 4r^2)^{1/2}\Bigg] + \nonumber \\
\frac{15}{2 \alpha \delta r}  \Bigg[3 q_0 q_2 \alpha^2 \delta^2 + 8  r^2 + 4 r (q_0 q_2 \alpha^2 \delta^2 + 4r^2)^{1/2} \Bigg] \times \nonumber \\
 \left \{  \left( q_0 q_2 \alpha^2 \delta^2 + 4r^2\right)^{1/2} + 2r \tanh \left[\frac{2r}{\alpha \delta} (\xi + \xi_0) \right]\right \} + \nonumber \\
\frac{15 q_2}{2} \left[\frac{3 }{r}(q_0 q_2 \alpha^2 \delta^2 + 4r^2)^{1/2} +2 \right]
\left \{  \left( q_0 q_2 \alpha^2 \delta^2 + 4r^2\right)^{1/2} + 2r \tanh \left[\frac{2r}{\alpha \delta} (\xi + \xi_0) \right]\right \}^2 + \nonumber \\
\frac{15 \alpha \delta q_2^2}{2 r}  \left \{  \left( q_0 q_2 \alpha^2 \delta^2 + 4r^2\right)^{1/2} + 2r \tanh \left[\frac{2r}{\alpha \delta} (\xi + \xi_0) \right]\right \}^3
\end{eqnarray}
Let us now consider another example for solution of equation of the class (\ref{class}).
In this example we set $m=2$ and $n=1$ in Eq.(\ref{class}). We shall assume $\Theta=2$ and thus from the balance equation $\nu = \Theta -1$ we obtain $\nu=1$. Then we shall search solution of the kind
\begin{equation}\label{y7}
u(\xi) = p_0 + p_1 g(\xi)
\end{equation}
where the differential equation of $g(\xi)$ is again the Riccati equation (\ref{y2}).
\par
The substitution of Eqs. (\ref{y7})and (\ref{y2}) in Eq.(\ref{class}) (where $m=2$ and
$n=1$) leads to the following equation

\begin{equation}\label{y8}
A_0 + A_1 g(\xi) + A_2 g(\xi)^2 + A_3 g(\xi)^3 + A_4 g(\xi)^4 =0
\end{equation}
Thus we obtain the following 5 nonlinear algebraic equations for the parameters of the equation
and parameters of the solution
\begin{eqnarray}\label{y9}
A_0&=& 3 \delta  q_2  +  \sigma  p_1  = 0
\nonumber \\
A_1&=& 12 \delta \alpha q_1  q_2 +  p_1  + 2 r q_2 + 2 \sigma \alpha p_0  q_2 + 3 \sigma \alpha p_1 q_1  = 0
\nonumber \\
A_2&=& \beta q_2 + \alpha p_0  q_2 + \alpha p_1 q_1 + 3 r \alpha  q_1 q_2 + 8 \delta \alpha^2 q_0  q_2^2 + 7 \delta \alpha^2 q_1^2  q_2 +\nonumber \\
&& 3 \sigma \alpha^2 p_0  q_1 q_2 + 2 \sigma \alpha^2 p_1 q_0 q_2 + \sigma \alpha^2 p_1 q_1^2 = 0
 \nonumber \\
A_3&=& \beta  q_1 + \alpha p_0  q_1 + \alpha p_1 q_0 + 2 r \alpha  q_0 q_2 + r \alpha  q_1^2 + 2 \sigma \alpha^2 p_0  q_0 q_2 + \sigma \alpha^2 p_0  q_1^2 + 
\nonumber \\ && \sigma \alpha^2 p_1 q_0 q_1 + 8 \delta \alpha^2 q_0  q_1 q_2 + \delta \alpha^2 q_1^3  = 0
\nonumber \\
A_4&=& \beta   + \sigma \alpha^2 p_0   q_1 + \alpha p_0   + 2 \delta \alpha^2 q_0  q_2 + \delta \alpha^2   q_1^2 + r \alpha  q_1 = 0
\nonumber \\
\end{eqnarray}
One solution of this system of equations is
\begin{eqnarray}\label{y10}
q_2 &=& - \frac{\sigma p_1}{3 \delta} \nonumber \\
q_1 &=& - \frac{2 \sigma^2 \alpha p_0 + 2 r \sigma - 3 \delta}{3 \alpha \delta \sigma} \nonumber \\
q_0&=& - \frac{2 \sigma^4 \alpha^2 p_0^2 + 4 r \sigma^3 \alpha p_0 + 2 r^2 \sigma^2 + 
	3 \delta r \sigma - 9 \delta^2 - 6 \alpha p_0 \sigma^2 \delta - 9 \beta \sigma^2 \delta}{6 p_1 \sigma^3 \delta \alpha^2} \nonumber \\
\end{eqnarray}
We shall use the solution (\ref{y4}) of the equation of Riccati (\ref{y2}). In this case
we must have 
$$
\pi = \frac{2 r \sigma - 2 \beta \sigma^2 - 3 \delta}{\delta \alpha^2 \sigma^2} >0
$$
When this relationship is fulfilled then the solution of the simplest equation (\ref{y2}) 
is
\begin{eqnarray}\label{y11}
g(\xi) = - \frac{2 \sigma^2 \alpha p_0 + 2 r \sigma -3 \delta}{2 \alpha \sigma^2 p_1} - \frac{3( 3\delta -2r\sigma + 2 \beta \sigma)}{2 \alpha^2 \sigma^3 p_1} \tanh \left[\frac{2 r \sigma - 2 \beta \sigma^2 - 3 \delta}{2 \delta \alpha^2 \sigma^2} (\xi + \xi_0) \right] \nonumber \\
\end{eqnarray}
The solution of the solved equation from
class (\ref{class}) ($m=2$, $n=1$) is
\begin{eqnarray}\label{y12}
u(\xi) = p_0 - p_1 \Bigg \{  \frac{2 \sigma^2 \alpha p_0 + 2 r \sigma -3 \delta}{2 \alpha \sigma^2 p_1} + \frac{3( 3\delta - 2 r\sigma + 2 \beta \sigma)}{2 \alpha^2 \sigma^3 p_1} \tanh \left[\frac{ 2 r \sigma -  2 \beta \sigma^2 - 3 \delta}{2 \delta \alpha^2 \sigma^2} (\xi + \xi_0) \right] \Bigg \} \nonumber \\
\end{eqnarray}
\section{Discussion and concluding remarks}
The study above has demonstrated the capability of the modified method of simplest equation for obtaining exact solutions of nonlinear partial differential equations connected to water waves theory.
Assuming vary large size of the system along the $x$-axis of the co-ordinate system we can impose
boundary conditions that can further fix the parameters of the solutions. Let us demonstrate this
on the basis of solution (\ref{y12}). We can impose, e.g. the following boundary conditions
\begin{equation}\label{y13}
u(\xi = \infty) = d_1; \ \ u(\xi=-\infty) = d_2
\end{equation}
where $d_1$ and $d_2$ are some appropriate constants. On the basis of Eq.(\ref{y12}) we arrive at the
system of equations
\begin{eqnarray}\label{y14}
 p_0 -   \frac{2 \sigma^2 \alpha p_0 + 2 r \sigma -3 \delta}{2 \alpha \sigma^2 } + \frac{3( \delta - 2r\sigma + 2\beta \sigma)}{ 2 \alpha^2 \sigma^3} = d_1 \nonumber \\
  p_0 -  \frac{2 \sigma^2 \alpha p_0 + 2 r \sigma -3 \delta}{2 \alpha \sigma^2} - \frac{3( \delta - 2 r\sigma + 2 \beta \sigma)}{2 \alpha^2 \sigma^3}  = d_2
\end{eqnarray}
From Eq.(\ref{y14}) we can determine two of the parameters, e.g, $\alpha$ and $\beta$. We obtain
\begin{eqnarray}\label{y15}
\alpha &=& \frac{3 \delta - 2 r \sigma}{\sigma^2[2 p_0 + p_1(d_1+d_2)]} \nonumber \\
\beta &=& \frac{1}{6 \sigma^3[2 p_0 + p_1(d_1+d_2)]^2}\Bigg[9 \delta^2  p_1 (d_1-d_2) - 36 \delta \sigma p_0^2 + 4 r^2 \sigma^2  p_1 (d_1 - d_2)  - \nonumber \\
&& 12 \delta r \sigma (d_1 - d_2) p_1 + 24 r \sigma^2 p_0^2 + 12 r \sigma^2 d_1 p_1^2 d_2 + 24 r \sigma^2 p_0 (d_1+d_2) p_1  - 18 \delta \sigma d_1 p_1^2 d_2 - \nonumber \\
&& 36 \delta \sigma p_0 (d_2 - d_1) p_1 + 6 r \sigma^2 (d_1^2 +  d_2^2) p_1^2   - 9 \delta \sigma (d_1^2 + d_2^2) p_1^2  \Bigg] \nonumber \\
\end{eqnarray}
Thus the solution (\ref{y12}) becomes
\begin{eqnarray}\label{y16}
u(\xi) &=& p_0 - p_1 \Bigg \{  \frac{2 \sigma^2 \alpha p_0 + 2 r \sigma -3 \delta}{2 \alpha \sigma^2 p_1} + \frac{3( 3\delta -2r\sigma + 2 \beta \sigma)}{2 \alpha^2 \sigma^3 p_1} \tanh \Bigg \{\frac{2 r \sigma - 2 \beta \sigma^2 - 3 \delta}{2 \delta \alpha^2 \sigma^2}\times
\nonumber \\
&& \Bigg[\frac{3 \delta - 2 r \sigma}{\sigma^2[2 p_0 + p_1(d_1+d_2)]} x + \frac{1}{6 \sigma^3[2 p_0 + p_1(d_1+d_2)]^2}\Bigg[9 \delta^2  p_1 (d_1-d_2) - \nonumber \\
&&36 \delta \sigma p_0^2 + 4 r^2 \sigma^2  p_1 (d_1 - d_2)  -  12 \delta r \sigma (d_1 - d_2) p_1 + 24 r \sigma^2 p_0^2 + 12 r \sigma^2 d_1 p_1^2 d_2 + \nonumber \\
&&24 r \sigma^2 p_0 (d_1+d_2) p_1  - 18 \delta \sigma d_1 p_1^2 d_2 -  36 \delta \sigma p_0 (d_2 - d_1) p_1 + \nonumber \\
&&6 r \sigma^2 (d_1^2 +  d_2^2) p_1^2   - 9 \delta \sigma (d_1^2 + d_2^2) p_1^2  \Bigg]t +\xi_0 \Bigg] \Bigg \} \Bigg \} \nonumber \\
\end{eqnarray}
This solution has many free parameters. This means that additional boundary conditions can be imposed,
e.g. $u'(x=-\infty)=0$; $u'(\infty)=0$, etc. We note that the solutions obtained
above are traveling waves of kink kind. One may search also for solitary waves
or for periodic waves. The results of this research will be reported elsewhere.

\end{document}